\documentclass[conference]{IEEEtran}
\IEEEoverridecommandlockouts
\usepackage{cite}
\usepackage{bbm}
\usepackage{float}
\usepackage{tikz}
\usepackage[english]{babel}
\usepackage{ifthen}
\usepackage[caption=false,font=footnotesize]{subfig}
\usepackage{graphicx}
\usepackage{hyphenat}
\usepackage{amsmath, amssymb,amsfonts}

\usepackage{algorithm}
\usepackage{algpseudocode}

\usepackage{multirow}
\usepackage[nohyperlinks]{acronym}
\usepackage[abbreviations]{foreign}  % For \ie, \eg, \etc
\usepackage{cleveref}
\usepackage[normalem]{ulem}
\usepackage{xcolor}
\usepackage{makecell}
\usepackage{framed}
\crefname{ALC@unique}{line}{lines}
\Crefname{ALC@unique}{Line}{Lines}

\def\BibTeX{{\rm B\kern-.05em{\sc i\kern-.025em b}\kern-.08em
    T\kern-.1667em\lower.7ex\hbox{E}\kern-.125emX}}

\makeatletter
\newcommand{\RNum}[1]{\uppercase\expandafter{\romannumeral #1\relax}}
\makeatother

\DeclareMathOperator*{\rgmin}{argmin}

\newcommand{\userset}{\mathcal{U}}
\newcommand{\dataset}[1]{\ratingmatrix_{#1}}
\newcommand{\itemset}{\mathcal{V}}
\newcommand{\ratingmatrix}{\mathcal{R}}
\newcommand{\model}{\Theta}
\newcommand{\argmin}[1]{\underset{#1}
{\rgmin}\;}
\newcommand{\loss}{\mathcal{L}^{rec}}
\newcommand{\embedding}[1]{e_{#1}}

\newcommand{\gradientp}[2]{\nabla \loss(\modelp{#2}; \dataset{#1})}
\newcommand{\lr}{\eta}
\newcommand{\modelp}[1]{\model^{#1}}
\newcommand{\modelu}[1]{\model_{#1}}
\newcommand{\modelpu}[2]{\modelp{#1}_{#2}}
\newcommand{\neighbors}[1]{N(#1)}
\newcommand{\neighborsp}[2]{\neighbors{#1}^{#2}}
\newcommand{\neighborsout}[1]{N_{out}(#1)}
\newcommand{\neighborsin}[1]{N_{in}(#1)}
\newcommand{\neighborsoutp}[2]{\neighborsout{#1}^{#2}}
\newcommand{\neighborsinp}[2]{\neighborsin{#1}^{#2}}
\newcommand{\edges}{\mathcal{E}}
\newcommand{\expectation}{\mathbb{E}}
\newcommand{\msgin}{M}
\newcommand{\Dtarget}{\itemset_{target}}
\newcommand{\momentumn}[1]{v_{#1}}
\newcommand{\momentum}[2]{v^{#1}_{#2}}

\begin{document}
\title{Inferring Communities of Interest in Collaborative Learning-based Recommender Systems\thanks{This work was supported by the French government grant managed by the Agence Nationale de la Recherche (ANR) through the France 2030 program with reference ``ANR-23-PEIA-005'' (REDEEM project), as well as the ANR Labcom program, reference ``ANR-21-LCV1-0012''.}
}
\author{
\IEEEauthorblockN{1\textsuperscript{st} Yacine Belal}
\IEEEauthorblockA{\textit{LIRIS, INSA Lyon} \\
Lyon, France \\
yacine.belal@insa-lyon.fr}
\and
\IEEEauthorblockN{2\textsuperscript{st} Mohamed Maouche}
\IEEEauthorblockA{\textit{INRIA, INSA Lyon, CITI, UR3720} \\
Villeurbanne, France \\
mohamed.maouche@inria.fr}
\and
\IEEEauthorblockN{3\textsuperscript{rd} Sonia Ben Mokhtar}
\IEEEauthorblockA{\textit{LIRIS, INSA Lyon, CNRS} \\
Lyon, France \\
sonia.benmokhtar@insa-lyon.fr}
\and
\IEEEauthorblockN{4\textsuperscript{th} Anthony Simonet-Boulogne}
\IEEEauthorblockA{\textit{iExec Blockchain Tech} \\
Lyon, France \\
anthony.simonet-boulogne@iex.ec}
}

\maketitle

\begin{center}
\textit{To appear in: Proceedings of the 45th IEEE International Conference on Distributed Computing Systems (ICDCS), 2025.}
\end{center}

\begin{abstract}
Collaborative-learning-based recommender systems, such as those employing Federated Learning (FL) and Gossip Learning (GL), allow users to train models while keeping their history of liked items on their devices. While these methods were seen as promising for enhancing privacy, recent research has shown that collaborative learning can be vulnerable to various privacy attacks. In this paper, we propose a novel attack called Community Inference Attack~(CIA), which enables an adversary to identify community members based on a set of target items. What sets CIA apart is its efficiency: it operates at low computational cost by eliminating the need for training surrogate models. Instead, it uses a comparison-based approach, inferring sensitive information by comparing users' models rather than targeting any specific individual model. To evaluate the effectiveness of CIA, we conduct experiments on three real-world recommendation datasets using two recommendation models under both Federated and Gossip-like settings. The results demonstrate that CIA can be up to 10 times more accurate than random guessing. Additionally, we evaluate two mitigation strategies: Differentially Private Stochastic Gradient Descent~(DP-SGD) and a \emph{Share less} policy, which involves sharing fewer, less sensitive model parameters. Our findings suggest that the \emph{Share less} strategy offers a better privacy-utility trade-off, especially in GL.
\end{abstract}


\section{Introduction}
Recommender systems are widely used algorithms in today's online services (\eg, movie recommendation in video-on-demand platforms, item recommendation in marketplaces, post recommendation in social media, \ldots etc.). Their usefulness in helping users dive into the overwhelming amount of content available online is no longer questionable. 
One of the downsides of today's recommender systems is their intrinsic centralization. Indeed, to compute useful recommendations, most recommender systems collect and process users' personal data such as their history of liked items, which may reveal sensitive information such as their age, sex, social status, health issues~\cite{friedman2015privacy}. This limitation has pushed the research community to investigate private recommender systems in the past decade~\cite{gao2020dplcf,rosinosky2021pprox}. 

More recently, in the aim of improving users' privacy, collaborative learning techniques have been adopted in recommender systems. As a result, several solutions have emerged with as a key common property the fact of collaboratively training a recommender machine learning model while keeping users' private data local. In this context, two families of recommender systems have emerged: (i) Federated Recommender Systems~\cite{liang2021fedrec++} (FedRecs) and (ii) Gossip Learning-based Recommender Systems~\cite{belal2022pepper} (GossipRecs). In FedRecs, a global model is trained locally by users across various rounds orchestrated by a central server. Only model updates are sent to this server. The latter aggregates the received models and sends the next round’s starting model to all or a subset of users. Instead, in GossipRecs, users train their own models locally and exchange them directly with a set of neighbors over a communication graph. 

However, due to the numerous attacks that have been proposed against collaborative learning systems (\eg,~\cite{yuan2023interaction,fu2022label}), a legitimate concern arises regarding the sensitivity of FedRecs and GossipRecs to privacy attacks. In this paper, we propose a novel attack called Community Inference Attack~(CIA). From an attacker perspective, the objective of CIA is to learn from the models received from clients, the $k$ users having the most similar set of recommended items to a given dataset crafted by the attacker from a catalog of available items. The novelty of CIA resides on the type of leakage and the resulting harm it may cause to users. Indeed, CIA allows an attacker to identify communities interested in a specific set of items (\eg, politically-oriented posts or videos, health-related subjects or places), which leaks sensitive information about them (\eg, minorities, protesters or people with a given disease). 

Various attacks have been proposed in the literature to leak private information from machine learning models. The closest attacks to CIA are membership inference attacks (MIAs)~\cite{DBLP:conf/sp/CarliniCN0TT22,yuan2023interaction} and attribute inference attacks (AIAs) ~\cite{chen2022practical,zhang2022comprehensive}. While the target of these attacks is different from the one of CIA, they could be leveraged as proxies to detect communities. In this paper we show that leveraging an MIA or an AIA to detect communities is both more costly and less effective than using CIA.

In this paper, we assess the leakage caused by a CIA on three recommender systems, one relying on Federated Learning~\cite{mcmahan2017communication} and two relying on Gossip Learning ~\cite{belal2022pepper, hegedHus2021decentralized}. We leverage two classical recommendation models, Generalized Matrix Factorization (GMF)~\cite{he2017neural} and Personalized Ranking Metric Embedding (PRME)~\cite{feng2015personalized}, to examine the impact of the recommendation model on privacy. These six configurations were tested on three real-life datasets. Besides evaluating the strength of the attack, we evaluate the impact of two countermeasures: Differentially Private Stochastic Gradient Descent (DP-SGD)~\cite{abadi2016deep} and the \emph{Share less} policy, which involves sharing a subset of model parameters~\cite{yuan2023interaction}.

Results show that in the Federated Learning setting, the attack can recover communities with an average accuracy exceeding random guessing by a factor of 10. Furthermore, in the Gossip Learning setting, the attack achieves an average accuracy three times better than random guessing. Augmenting GossipRecs with DP-SGD and the \emph{Share less} policy enhances their resilience to CIA, with \emph{Share less} offering a better privacy-utility trade-off than DP. 

\noindent Our contributions are as follows:
\begin{itemize}
\item We propose a novel attack aimed at discovering communities based on their preferences in collaborative-learning-based recommender systems, serving as a new privacy metric to evaluate a new type of leakage from model exchanges in distributed settings.
\item We present this attack both in FL and Gossip settings. While the attacker in FL is assumed to be placed on the server side, in the Gossip setting we ran experiments considering all possible attacker placements in the communication graph. We further considered a proportion of colluding nodes in this setting.
\item We evaluate the leakage from our attack across different recommender models, using three real-life datasets with approximately 700 to 1000 users each.
\item We compare the privacy-utility trade-off of two mitigation mechanisms, DP-SGD and the \emph{Share less} policy.
\item We investigate how other attacks in the literature namely MIA and AIA can be leveraged as proxies for CIA and evaluate their efficiency in detecting communities.
\end{itemize}

The rest of the paper is organized as follows: Section~\ref{sec:illustration} serves as a concrete illustration of the impact of CIA when targeting health vulnerable users in the Foursquare dataset. Section~\ref{sec:background} provides background on Collaborative-Learning Recommender Systems and the defense mechanisms: the \emph{Share less} strategy and DP-SGD. Section~\ref{sec:CIAoverview} details our threat model and the attack algorithm. Section~\ref{sec:experiments} describes our evaluation setting, followed by an evaluation of the attack (Section~\ref{sec:attackeval}) and defense mechanisms (Section~\ref{sec:defenseeval}). We then analyze CIA's sensitivity to various parameters while comparing it with MIAs and AIAs (Section~\ref{subsec:sensitivity}) and discuss our design choices (Section~\ref{sec:discussion}). Finally, we review related work in Section~\ref{sec:relatedwork} and conclude in Section~\ref{sec:conclusion}.

\section{Motivating Example\label{sec:illustration}} 
\begin{figure}[H]
    \centering
    \includegraphics[width=0.45\textwidth]{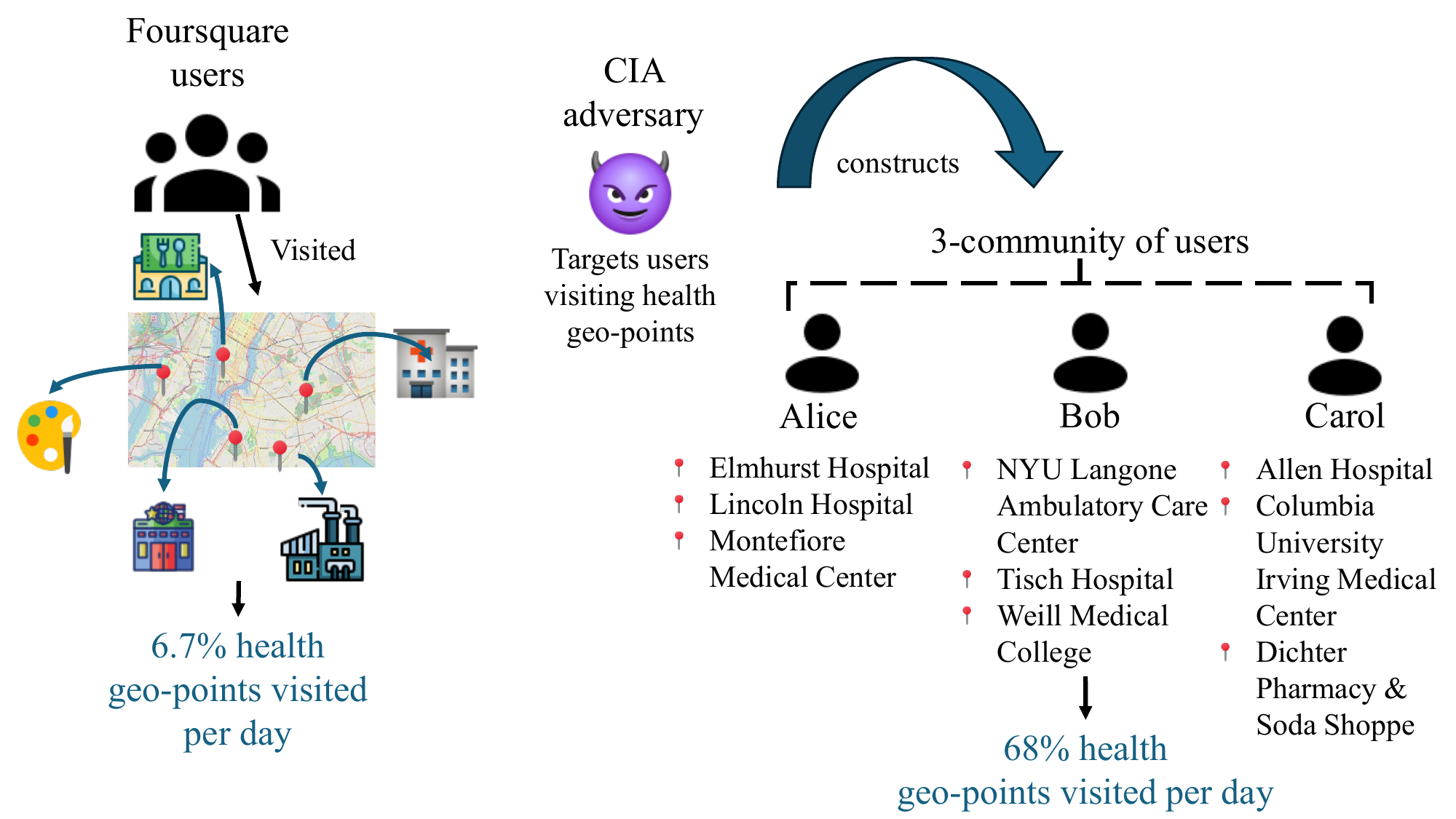}
    \caption{CIA run targeting "health vulnerable" users in the Foursquare dataset.}
    \label{fig:illustration}
\end{figure}

Consider the Foursquare dataset, which consists of timestamped geo-localization points visited by users. Each point in this dataset can be categorized semantically (\eg, Health and Medicine, Retail\ldots etc.) , as detailed in a previous work~\cite{wiedemann2024you}.
In order to illustrate the risks incurred by collaborative learning-based recommender systems, we performed the following experiment. We implemented a point-of-interest recommender system on the aforementioned dataset. The recommender system is trained over Federated Learning. We ran the attack described in this paper on the server side with the objective of identifying "health vulnerable" users (\ie, those who frequently visit health categorized geo-points). As illustrated in Figure~\ref{fig:illustration}, by leveraging our proposed attack and utilizing only the users' models and publicly available categorizations, the adversary successfully identifies a community of three users. These users are characterized by having at least 68\% of health geo-points visits per day compared to only 6.7\% for the overall Foursquare users. Moreover, by using a fine-grained set of geo-points when running CIA, the adversary can achieve a relatively accurate estimate of the specific health points visited by the community (\eg, Elmhurst Hospital). This example highlights the potential privacy risks inherent to the notion of community, such as discrimination by insurers and targeted exploitation through personalized health-related advertisements.

\section{Background}\label{sec:background}
In this section we present the necessary background on recommender systems (Section~\ref{subsec:recsys}) with a specific focus on Federated Learning-based Recommender Systems (Section~\ref{subsec:fedrecs}) and on Gossip Learning-based Recommender Systems (Section~\ref{subsec:gossiprecs}). We then describe the defense mechanisms, namely, Share less (Section~\ref{subsec:shareless}) and DP-SGD~(Section~\ref{subsec:dp}).

\subsection{Recommender Systems}\label{subsec:recsys}
Let $\ratingmatrix \in \mathbb{R}^{|U| \times |V|}$ be an interaction (\eg, rating) matrix for a set of users (clients) $\userset$ and a set of items $\itemset$. In this work, we are interested in Collaborative Filtering~(CF) recommenders, which aim at capturing
the relationship between $\mathcal{U}$ and $\mathcal{V}$ by abstracting each user $i \in \userset$ and item $j \in \itemset$ into real-valued vectors (embeddings), $\embedding{i} \in E^{\userset}$ and $\embedding{j} \in E^{\itemset}$, respectively. This set of embeddings $E = \{E^{\userset},E^{\itemset}\}$ is part of a model $\model$ ($E \subset \model)$, which is learned by solving an optimization problem of the form:
\begin{equation}
    \argmin{\model} \loss(\model; R)
\end{equation} where $\loss$ is a loss function. In our settings, Stochastic Gradient Descent~(SGD) is leveraged to solve this problem.

\subsection{Federated Recommender Systems}\label{subsec:fedrecs}
Federated Recommender Systems~(FedRecs) allow multiple users to collaboratively train a recommender model while keeping their data on their premises. The training process consists of several communication rounds between clients, and a central server. Specifically, at each round $t \in [T]$, (1) (selected) users receive the global model $\modelp{t}$ from the server; (2) they run several training steps using their local data $\dataset{u}$, obtaining local model updates (gradients), say $\gradientp{u}{t}$; (3) they send back the updates to the server. To form the new global model, the server aggregates them as in  $\modelp{t+1} = \modelp{t} - \lr \sum_{u \in \userset} W_u \gradientp{u}{t}$, where $\lr$ is a learning rate and $W$ is a weighting vector that depends on the aggregation rule. Finally, this process is repeated until model convergence. 

% While FedRecs are generally regarded as privacy-aware, they do not provide absolute guarantees of privacy. In fact, their privacy vulnerabilities are even more pronounced in CF models due to the discriminatory latent features present in the embeddings. This creates a unique privacy risk that needs to be quantified in the context of FedRecs.
% Therefore, it is critical to quantify the specific risks posed by these models and assess their impact on federated learning. In the next section we present a gossip-based approach to relax the single server assumption.
% 
\subsection{Gossip-based Recommender Systems}\label{subsec:gossiprecs}
Gossip-based Recommender Systems~(GossipRecs) are used to train a recommendation model across a network of users connected in a peer-to-peer fashion. In this setting, each user $u$ maintains a local model $\modelu{u}$, while the (dynamic) communication network is often modeled by a set of directed graphs $G^1, \ldots G^T$ where $G^t = (\userset, \edges^t)$ represents the network at round $t$ and $\edges^t \in \userset^2$ is the set of 
communication links (\ie, edges) between users. Let $\neighborsp{u}{t} = \{\neighborsinp{u}{t} \bigcup \neighborsoutp{u}{t} \}$ denote the set of neighbors of $u$ (view) at round $t$. Similarly to other works~\cite{mrini2024privacy,pasquini2022privacy}, we consider  P-out-regular graphs (\ie, $\expectation(|\neighborsin{u}|) = P$ and $|\neighborsout{u}| = P$). 

At each iteration $t$, (1) a random user  $u$ (or subset of users) wakes up and casts $\modelpu{t}{u}$ to a randomly sampled neighbor $v \in \neighborsoutp{u}{t}$. In a second phase, (2) $u$ aggregates the models received since the last waking up step, say 
$\msgin^{<t}_{in}(u)$, as in $\modelpu{t+\frac{1}{2}}{u} = W_{(u,\modelpu{t}{u})} \modelpu{t}{u} + \sum_{\model \in \msgin^{<t}_{in}(i)}  W_{(u,\model)} \model$. where $W_{(u,m)}$ denotes the weight user $u$ assigns to model $m$ in the aggregation. Finally, (3) $u$ forms its new model $\modelpu{u}{t+1}$ by making local training steps. Furthermore, users periodically explore new neighborhoods through a random peer-sampling protocol~\cite{busnel2011uniformity}. 
% We further discuss our setting choices in~\cref{subsec:choice}.
\subsection{Share less strategy}\label{subsec:shareless}
User embeddings encode latent features describing users' semantics, making them more vulnerable. Based on this rationale, Yuan et al.~\cite{yuan2023interaction} have advocated for retaining  $\embedding{u}$, $\forall u \in \userset$, on the user $u$'s device ($E^{\userset}$ is private), while regularizing item embeddings updates' to reduce their sensitivity (See Equation~\ref{eq:shareless}). This approach has been shown to improve the resilience against MIAs. We quantify its effectiveness against CIA.

\begin{equation}\label{eq:shareless}
    \mathcal{L} = \mathcal{L}^{rec} + \tau \sum_{j \in \itemset_{u}} ||\embedding{ju}^t - \embedding{j}^t||_2
\end{equation}

where $\embedding{j}$ denotes the global embedding of item $\embedding{j}$ and $\embedding{ju}$ the updated embedding of item j at user $u$, while $\tau$ is the regularization factor. In GL, where there are no global embeddings, $\embedding{j}^t$ is replaced with $\embedding{ju}^{t-1}$. 

%  While~\cite{yuan2023interaction} showcased improved resilience against Membership Inference Attacks (MIA), the effectiveness of this strategy against CIA, which infers sensitive information by comparing models, remains unclear. Additionally, the impact of this mechanism on utility, especially in an asynchronous Gossip setting, has not been evaluated.

\subsection{Differentially Private SGD}\label{subsec:dp}
Differentially Private SGD~(DP-SGD)~\cite{abadi2016deep} is an optimization algorithm that trains a model $\model$ while ensuring that neighboring input datasets (that differ in one datapoint) produce model updates with similar statistical properties (up to an $\epsilon$). To this end, gradients are clipped by a predetermined threshold $C$ and a calibrated noise, drawn from some probability distribution, is added to them during the SGD process.  This has the effect of bounding the contribution of (batches of) data points. In our case, we opt for the Gaussian Distribution (\ie, noise $\sim \mathcal{N}(0, {(\iota C)}^2 I)$ , where $\iota$ is a scaling factor). Moreover, we leverage a version of DP, referred to as local DP~(LDP), where noise is added at user-level. 

While DP-SGD provides formal privacy guarantees, it often incurs a substantial utility cost~\cite{bagdasaryan2019differential}. One of the objective of this work is therefore to investigate the privacy-utility trade-off of LDP in the context of CIA.
\section{Community Inference Attack~(CIA)}~\label{sec:CIAoverview}
\begin{figure}[h!]
\centering
\includegraphics[width=\linewidth]{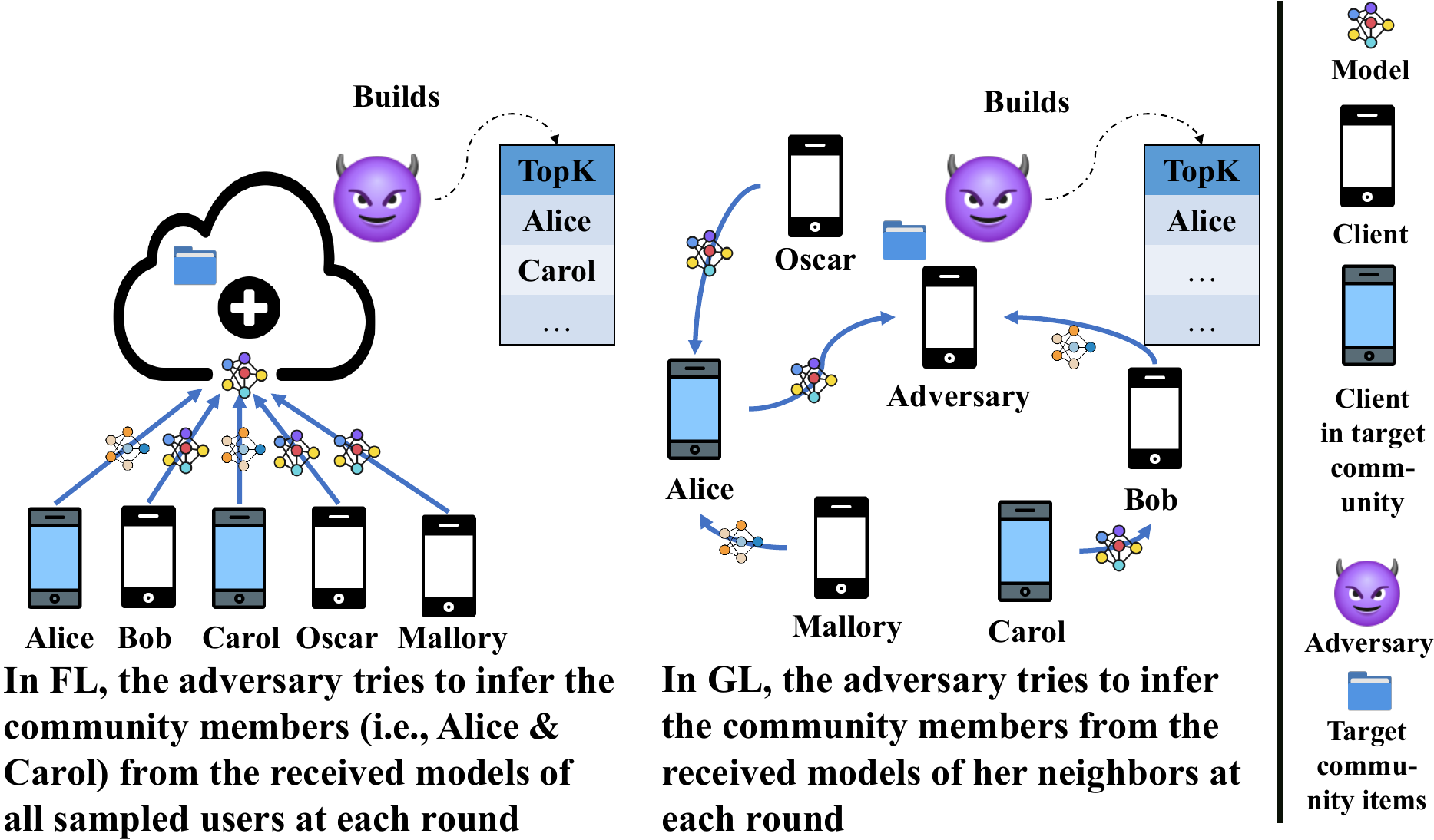}
\caption{Community Inference Attack (CIA) : a comparison between FL and GL settings.\label{fig:CIA_overview}}
\end{figure}

We present in this section our threat model and assumptions  (Section~\ref{subsec:threatmodel}) before delving into our Community Inference Attack~(CIA) (Section~\ref{subsec:algo}).

\subsection{Threat Model and Adversary's Knowledge}\label{subsec:threatmodel}
We assume $N$ users involved in a collaborative learning algorithm of a joint recommendation model $\model$. We assume the presence of an honest-but-curious adversary $A$. That is to say, the adversary complies with the learning protocol while aiming to learn sensitive information. We assume that $A$ has access to a set of items, $\Dtarget$, that are representative of the community to be found. In real life, $\Dtarget$ can be handcrafted from a catalog of available items as these are usually publicly available as we did in the illustrating example (Section~\ref{sec:illustration}). The primary objective for the adversary is to identify a community of users, denoted as $C$ which exhibits the closest item preferences to the target item set $\Dtarget$. In this work, we distinguish the following three settings:\\ 
\textbf{Federated setting:} In this setting (Figure~\ref{fig:CIA_overview}, Left), we assume that the adversary takes control of the federated server. The server may contact all or part of the users in each FL round. It then uses the received models to carry out the attack.\\
\noindent\textbf{Gossip-based setting with a single adversary:} We first consider an adversary with a single participant under its control (Figure~\ref{fig:CIA_overview}, Right). This adversary carries out her attack by using the models she receives from her neighbors throughout the execution of the protocol. Note that as we rely on a random peer sampling protocol, the adversary will periodically have new neighbors from whom she will receive models. Moreover, considering our P-out regular-graphs, the expected number of models observed by the adversary is equal to P. That is, $\mathbb{E}[$Adv. seen models$] = P$.\\
\noindent\textbf{Gossip-based setting with an adversary controlling several nodes:} In this setting, we consider an adversary that controls a proportion of $f/N$ participants that are randomly positioned in the network. As we assume the attacker to follow the honest-but-curious threat model, these participants, referred to as colluders, only share their knowledge with each other (~\cref{lst:line:colluders} in~\cref{alg:att_gl}). Consequently, in this setting, $\mathbb{E}[$Adv. seen models$] \le f \cdot P$.

\subsection{Adversary Goal and CIA's Algorithm}\label{subsec:algo}
The objective of the attacker is to discover a community $C$ of $K$ users that are the most likely to be interested in the set of items $\Dtarget$. Towards this purpose, $A$ uses models received from participants during the learning process (See~\cref{lst:line:receive,lst:line:receive2} in \cref{alg:att_fl,alg:att_gl} for FL and GL, respectively).
Specifically, the attack exploits the fact that a model is likely to assign higher relevance scores for items rated by users coming from the same community.
To capitalize on this, an adversary can compute the relevance score assigned by each received model $\modelu{u}$ to $\Dtarget$ (See~\cref{lst:line:attack,lst:line:attack2} of \cref{alg:att_fl,alg:att_gl}, respectively). The adversary can then rank the models over time, obtaining ~$\hat{C}$, the set of $K$ users most likely to belong to the community $C$ (~\cref{lst:line:return,lst:line:return2} of \cref{alg:att_fl,alg:att_gl}, respectively). This can be written as follows:
\begin{equation}
\label{eq:Au}
\begin{split}
    \hat{C} &= \{ 
    c_1, c_2,\ldots, c_K \}
    \quad \text{such that} \quad
    \forall k \leq K,\forall u \in \userset \smallsetminus \hat{C}:\\
    &\hat{Y}(\modelu{c_k}, \Dtarget) \geq \hat{Y}(\modelu{u}, \Dtarget)
\end{split}
\end{equation}

Where $\hat{Y}(\modelu{u}, \Dtarget) = \frac{1}{|\Dtarget|} \sum_{i \in \Dtarget} \hat{y}_{ui}$ and ${\hat{y}}_{ui}$ is the relevance score assigned by $\modelu{u}$ to item $i$. In practice, the relevance can be any recommendation quality metric. 

\begin{algorithm}[htbp]
\caption{CIA Algorithm run by adversary $A$ in FL}
\label{alg:att_fl}
\begin{algorithmic}[1]
\Require $\Dtarget$: target dataset, $K$: size of the community, $\beta$: momentum coefficient
\Ensure $\hat{C}$: predicted community

\ForAll{$u \in \userset$}
    \State $\momentumn{u} \gets \text{Dict()}$ \label{lst:line:init}
\EndFor

\State $\hat{C} \gets \{\}$

\For{$t = 1$ to $T$}
    \ForAll{$u \in \userset$} \label{lst:line:receive}
        \If{$t \ne 0$}
            \State $\momentumn{u} \gets \beta \cdot \momentumn{u} + (1 - \beta) \cdot \modelu{u}$ \label{lst:line:momentum}
        \Else
            \State $\momentumn{u} \gets \modelu{u}$ \label{lst:line:momentuminit}
        \EndIf
        \State $y_u \gets \text{EvaluateModel}(\momentumn{u}, \Dtarget)$ \label{lst:line:attack}
        \State $\text{AddSorted}(\hat{C}, u, y_u)$
    \EndFor
\EndFor

\State $\text{Slice}(\hat{C}, K)$ \label{lst:line:return}

\end{algorithmic}
\end{algorithm}

\begin{algorithm}[htbp]
\caption{CIA Algorithm run by adversary $A$ in GL}
\label{alg:att_gl}
\begin{algorithmic}[1]
\Require $\Dtarget$: target dataset, $K$: size of the community, $\beta$: momentum coefficient, $A_{colluders}$: Colluding partners
\Ensure $\hat{C}$: predicted community

\ForAll{$u \in \userset$}
    \State $\momentumn{u} \gets \text{Dict()}$ \label{lst:line:init2}
\EndFor

\State $\hat{C} \gets \{\}$

\While{Training}
    \ForAll{$\modelu{u} \in \msgin_{in}^{< t}(A)$} \label{lst:line:receive2}
        \If{$t \ne 0$}
            \State $\momentumn{u} \gets \beta \cdot \momentumn{u} + (1 - \beta) \cdot \modelu{u}$ \label{lst:line:momentum2}
        \Else
            \State $\momentumn{u} \gets \modelu{u}$ \label{lst:line:momentuminit2}
        \EndIf
        \State $y_u \gets \text{EvaluateModel}(\momentumn{u}, \Dtarget)$ \label{lst:line:attack2}
        \State $\text{AddSorted}(\hat{C}, u, y_u)$
        \State $\text{Multicast}(A_{colluders}, u, \momentumn{u})$ \label{lst:line:colluders}
    \EndFor
\EndWhile

\State $\text{Slice}(\hat{C}, K)$ \label{lst:line:return2}

\end{algorithmic}
\end{algorithm}

\subsubsection{\textbf{FL vs GL}} The only difference between FL and GL is that the attacker does not encounter participants in the same manner (See Section~\ref{subsec:threatmodel}). While  in  FL, (the server acting as an adversary) receives  models from  the participants of each FL round (line~\ref{lst:line:receive} in Algorithm~\ref{alg:att_fl}), in GL, adversaries receive models from their neighbors or colluders (line~\ref{lst:line:receive2} in Algorithm~\ref{alg:att_gl}). Besides this difference, the attacker's actions remain the same. 

\subsubsection{\textbf{Colluders in GL}} When the attacker controls more than one node in GL, she forwards models received from her neighbors to her colluders (line~\ref{lst:line:colluders} in Algorithm \ref{alg:att_gl}).

\subsubsection{\textbf{Limiting the Effect of Model Aging through Momentum}}
Models tend to leak more information on their training data in early stages of training. Furthermore, in GL, models arrive at varying stages of the learning process, depending on the adversary's neighborhood (\ie, temporality). This makes model comparison even more difficult because the relevance score could be high due to the overall quality of the model rather than its specification to $\Dtarget$. To account for these phenomena, we implemented a momentum technique. More specifically, instead of computing relevant scores directly on the new received model $\modelpu{t}{u}$, we use $\momentum{t}{u}$, an aggregated model through rounds (See Equation~\ref{eq:momentum} and \cref{lst:line:momentum,lst:line:momentum2} of \cref{alg:att_fl,alg:att_gl}).

\begin{equation}
\label{eq:momentum}
\momentum{t}{u} = \beta \times \momentum{t-1}{u} + (1 - \beta) \times \modelpu{t}{u}   
\end{equation}
Where $\beta$ is the momentum coefficient and $ \momentum{0}{u} = \modelpu{0}{u}$ (\cref{lst:line:momentuminit,lst:line:momentuminit2} in~\cref{alg:att_fl,alg:att_gl}). %\forall u \in \mathcal{U},

\subsection{\textbf{Adapting CIA to the Share less strategy}}
As discussed previously, an adversary in CIA needs to compute the relevance score $\hat{Y}(\modelu{u}, \Dtarget)$. However, this cannot be achieved straightforwardly in the Share less strategy, where the adversary only receives partial models (\ie, $E^{\userset}$ is private). To account for this, we propose a methodology that allows the adversary to obtain a sufficiently accurate estimate of the relevance score. More precisely, the adversary creates a fictional interaction matrix $\dataset{A}$ using the target set of items $\Dtarget$ in order to train a user embedding  $\embedding{A}$. Semantically, this embedding represents a fictive user that likes items of $\Dtarget$. Then, upon each reception of a partial model (from a victim) in the form $E^{\itemset}$ (and possibly non-embedding parameters), the adversary combines it with $\embedding{A}$ to form an approximation of the model to be attacked, say $\hat{\model}_{u}$. Finally, $A$ can conduct CIA by replacing $\modelu{u}$ with $\hat{\model}_{u}$ in~\cref{lst:line:attack,lst:line:attack2} of~\cref{alg:att_fl,alg:att_gl}. This approach works because of the comparative nature of CIA, as learning one user embedding provides a reference basis to compare items embeddings coming from different users.

% In this work, we empirically assess the extent of utility loss associated with this algorithm in the context of FedRecs and GossipRecs (See Sections~\ref{subsec:Q7} and~\ref{subsec:Q8}). 

\section{Experimental Setup}
\label{sec:experiments}
We present in this section our experimental setup. First, we describe the datasets and recommendation models used. Then, we define our evaluation metrics (Section~\ref{subsec:metrics}). Finally, we describe the evaluated protocols and baselines (Section~\ref{subsec:protocols}). 

\subsection{Datasets}
\label{subsec:datasets}
\begin{table}
\centering
\caption{Summary of Datasets.\label{tab:datasets}}
\tabcolsep=0.11cm
\begin{tabular}{|c|c|c|c|}
\hline
\textbf{Dataset} & \textbf{Users} & \textbf{Items} & \textbf{Ratings} \\ \hline
MovieLens-100k & 943 & 1682 & 100k movie ratings \\
Foursquare-NYC~\cite{yang2014modeling} & 1083 & 38333 & 200k check-ins \\
Gowalla-NYC~\cite{chofriendship} & 718 & 32924 & 185,932 check-ins \\
\hline
\end{tabular}
\end{table}
²
We used the datasets described in Table~\ref{tab:datasets}. Note that we
pre-process these datasets in the following way: we assign
the value one to all user ratings/check-ins (i.e., observed
interactions) and the value zero to non-observed ones, as is classically done for 
classification-based recommendation~\cite{he2017neural}.

\subsection{Recommendation Models and Hyper-Parameters}
In this work, we leveraged two standard recommendation models: Generalized Matrix Factorization~\cite{he2017neural} and Personalized Ranking Metric Embedding~(PRME)~\cite{feng2015personalized}. For both models, we utilize the same hyper parameters as the original works. For the gossip algorithms, we set the out-neighborhood size to 3 and the periodic change of views follows $p \sim Exp(0.1)$, while for the personalized gossip algorithm, we set the exploration ratio to 0.4. Finally,
unless stated differently, the momentum coefficient $\beta = 0.99$ for all settings.

% \begin{table}
% \centering
% \caption{Default learning parameters. Note that \text{Exp}(0.1) is the exponential law with scale 0.1.}\label{tab:hyperparameters}
% \begin{tabular}{lll}
% \hline
% \textbf{Parameter} & \textbf{GMF}     & \textbf{PRME}   \\ \hline
% Embedding size        & 8    & 60      \\
% Initialization                                       & $\mathcal{N}(0,0.01)$ & $\mathcal{N}(0,0.01)$\\
% Optimizer             & Adam~\cite{kingma2014adam} & SGD     \\
% Learning rate         & 0.01 & 0.005   \\
% Batch Size            & 64   & 128 \\
% Regularization term   & 0.001    & 0.03    \\
% Time window threshold & NA   & 6 hours \\
% Component weight      & NA   & 0.2     \\ 
% Gossip peer-sampling periodicity & \multicolumn{1}{c}{\text{Exp}(0.1)} \\
% Pers-gossip exploration ratio &  \multicolumn{1}{c}{0.4}  \\
% \hline
% \end{tabular}
% \end{table}

\subsection{Evaluation Metrics}
\label{subsec:metrics}
In this section, we present the evaluation metrics and gossip protocols that were employed in our experimental evaluation. 

\noindent \textbf{Ground truth:} To measure the performance of CIA, the following ground truth was considered: for a given set of items $\Dtarget$, the true community $C$ is defined as the top-K users whose training sets of items are most similar to $\Dtarget$. Due to the nature of the data (\ie, sets), similarity is measured using the Jaccard index~(see \cref{eq:jaccard}). Moreover, to ensure the statistical significance of the results, we made each user $u$ (separately) play the role of the adversary by using $\itemset_{train}^{(u)}$ as $\Dtarget$. 

% \yacine{next sentence could be clarified}the role of $\Dtarget$ is played by each training set of items for each user in $\userset$.

% \yacine{here}To extensively assess our attack, in all our experiments, we made each user $u$ (separately) play the role of the adversary by using her own train set as the target set $\Dtarget$. Her objective is then to find her $K$ most similar users, which constitutes the community $C$. Since we have binary rating data, we use the Jaccard similarity measure~\cite{ivchenko1998jaccard} to find a community $C$ composed of users who have the most items in common with $u$ (see \cref{eq:jaccard}). Other similarity metrics could be used (\eg. cosine, sorensen-dice)).

\begin{equation}
\label{eq:jaccard}
\begin{split}
&\mathcal{C} = \{ c_1, c_2,\ldots, c_K \} 
\quad \text{such that} \quad \forall k \leq K, \forall u \in \mathcal{U} \smallsetminus C , \\ &\quad  \Biggl \lvert \frac{\Dtarget \bigcap \itemset_{train}^{(c_k)}} {\Dtarget \bigcup \itemset_{train}^{(c_k)}} \Biggr \rvert \geq \Biggl \lvert \frac{\Dtarget \bigcap \itemset_{train}^{(u)}} {\Dtarget \bigcup \itemset_{train}^{(u)}} \Biggr \rvert 
\end{split}
\end{equation}

\noindent where $\itemset^{(u)}_{train}$ is the set of items used in training by user $u$ while $K$ is a parameter defining the size of user communities. Unless stated otherwise, we set K=50.

\noindent \textbf{Attack Accuracy (at round R):} Given $\Dtarget$ and the inferred community $\hat{C}$, the accuracy at rank K is defined as:
\begin{equation}
Accuracy@R(\Dtarget) = \frac{\lvert \hat{C}_{target} \bigcap C_{target} \lvert}{K} 
\end{equation}

\noindent \textbf{Average Attack Accuracy (at round R):} It is defined as the average accuracy over all possible $\Dtarget$ (adversaries).

\noindent \textbf{Maximum Attack Accuracy (over all rounds):} It is the maximum of average attack accuracies. We refer to it as Max AAC. 

\noindent \textbf{Best 10\% AAC:} It refers to the minimum attack accuracy achieved by the best 10\% attackers at the round where Max AAC is obtained. 

\noindent \textbf{Accuracy upper bound:} It is defined as the maximum accuracy that an adversary could achieve considering the models observed. For instance, an adversary targeting $C$ and having interacted with $p * |C|$ proportion of community users has $Accuracy \le p$. This bound is equal to 1 for the FL settings.
\noindent \textbf{Recommendation quality metrics:} As utility metrics, we leverage the hit ratio (HR) at rank K~\cite{he2017neural} to evaluate GMF and the F1-Score For PRME.

% \noindent In addition to these metrics, we also report Cumulative Distribution Functions~(CDFs) to investigate individual attack accuracy (of all possible attackers). This allows us to quantify lower and upper bounds for attack accuracy, and can be observed in Figure~\ref{fig:cdf_shareless} for the attack evaluation and Figure~\ref{fig:cdf_shareless} for the Share less strategy evaluation.

\subsection{Evaluated Protocols and Baselines}\label{subsec:protocols}
We implemented the following protocols:\\
\noindent \textbf{FL:} The classical FedAvg~\cite{mcmahan2017communication}  algorithm. 

\noindent \textbf{Rand-Gossip~\cite{hegedHus2021decentralized}:} A standard adaptation of FedAvg in a decentralized setting with a random peer-sampling. 

\noindent \textbf{Pers-Gossip~\cite{belal2022pepper}:} A personalization oriented GL protocol where the performance of models is considered during peer-sampling. This version allows us to study the impact of personalization on CIA. 

\noindent \textbf{Random guess:} It is defined as a random draw of K elements from N without replacement, which follows a hyper-geometric law $\mathcal{G}(K,K,N)$. Therefore, the expected value of a random guess is $\frac{K}{N}$.

\section{Attack Evaluation}\label{sec:attackeval}
In this section, we investigate the impact of CIA on the three settings described in Section~\ref{subsec:threatmodel}. More specifically, we aim at answering the following research questions:
\begin{itemize}
%\item \textbf{RQ1}: What is the impact of CIA on FedRecs?
\item \textbf{RQ1}: How vulnerable are FedRecs to CIA?
\item \textbf{RQ2}: How vulnerable are GossipRecs to CIA?
%\item \textbf{RQ2}: What is the impact of CIA on GossipRecs compared to FedRecs?
\item \textbf{RQ3}: What is the impact of GossipRecs personalization schema on the accuracy of CIA?
%\item \textbf{RQ3}: What is the impact of GossipRec personalization schema on the accuracy of CIA and how does it compare with FedRecs?
\item \textbf{RQ4}: What is the impact of colluders in GossipRecs on the performance of CIA?
\end{itemize}

We answer these questions in sections~\ref{subsec:Q1}, \ref{subsec:Q2}, \ref{subsec:Q3}, and~\ref{subsec:Q4}, respectively.

\subsection{CIA on FedRecs}\label{subsec:Q1}
\begin{table}
\centering
\caption{Summary of Attack results in the Federated Setting. Accuracy upper bound over all configurations is 100\%.\label{tab:fl}}
\tabcolsep=0.11cm
\begin{tabular}{|c|c|cc|} 
\hline 
\multirow{2}{*}{\begin{tabular}[c]{@{}c@{}}Dataset\\ (Random Bound)\end{tabular}} & \multirow{2}{*}{Model} & \multicolumn{2}{c|}{Metric}                                       \\ \cline{3-4} &                        & \multicolumn{1}{c|}{Max AAC \%} & \multicolumn{1}{l|}{Best 10\% AAC \%} \\ \hline \multirow{2}{*}{\begin{tabular}[c]{@{}c@{}}Foursquare\\ (3.6)\end{tabular}}       & GMF                    & \multicolumn{1}{c|}{44.93}   & 66                                 \\ \cline{2-4} & PRME                   & \multicolumn{1}{c|}{18.4}    & 34                                 \\ \hline \multirow{2}{*}{\begin{tabular}[c]{@{}c@{}}Gowalla\\ (5)\end{tabular}}            & GMF                    & \multicolumn{1}{c|}{57.3}    & 77                                 \\ \cline{2-4} & PRME                   & \multicolumn{1}{c|}{32}      & 52                                 \\ \hline \begin{tabular}[c]{@{}c@{}}Movielens\\ (5.3)
\end{tabular}                         & GMF                    & \multicolumn{1}{c|}{57.4}    & 76                                 \\ \hline \end{tabular}
\end{table}

In Table~\ref{tab:fl}, we depict the results of CIA on FedRecs. The results depicted in this table correspond to measures performed by launching as many experiments as there are users in the corresponding dataset, where each time the FL server uses the train set of a single user as $\Dtarget$.

We observe that CIA obtains a high privacy leakage. Especially, when using GMF, since the server manages to reach between 44.93\% and 57.4\% average accuracy (depending on the dataset). This is more than 10 times the random bound (3.6\%-5.3\%). Naturally, we observe an even larger privacy leakage achieved by the best 10\% attackers, which, for instance, achieve at least 77\% attack accuracy on Gowalla with PRME. Notably, PRME seems less sensitive to the attack (between 18.4\% and 32\%). The reason behind this is that PRME is designed to learn a more challenging task compared to GMF. This is further illustrated by its lower utility (see Figure~\ref{fig:tradeoff_prme}).
% From a distribution perspective, Figure~\ref{fig:cdf_shareless} confirms that individual attack accuracies are
% consistent with the average accuracy. Specifically, we observe that at least 90\% of adversaries achieve a higher accuracy than the random guess. These findings are further supported with the PRME-G model where 75\% and 61\% of adversaries surpass the random guess on Foursquare and Gowalla,
% respectively. On the whole, except for the PRME-G model on Foursquare, a majority of potential adversaries
% (75\%-90\%) achieve higher accuracy than the random guess and at least 35\% achieve a higher accuracy than the
% reported average accuracy. 
Overall, an adversary clearly manages to leverage received models to find communities of participants in FedRecs~\textbf{(RQ1)}.

\subsection{CIA on GossipRecs}\label{subsec:Q2}
Compared to the results of FedRecs, Table~\ref{tab:gl} illustrates a reduced Max AAC in GossipRecs compared to FedRecs. Specifically, the highest accuracies drop from 57.3\% to 12.2\% and from 57.4\% to 14.6\% for Gowalla and MovieLens (respectively) with the GMF model. Nevertheless, these accuracies are still at least twice as large as the random guess. This indicates that GossipRecs are less vulnerable to CIA but there still is a significant privacy leakage compared to the random bound~\textbf{(RQ2)}. 

%\mohamed{dans la suite on compare bcp FedRecs et GossipRecs, est ce le but ?}
% From a distribution perspective, Figure~\ref{fig:cdf_shareless} allows us to observe that the gap between GossipRecs and FedRecs is even more pronounced when considering individual attack accuracies. Specifically, the top 1\%-5\% adversaries do not achieve over 60\% of accuracy over all datasets in GossipRecs, whereas they would achieve up to 100\% in FL. Furthermore, we observe that for Rand-Gossip, 81.56\%, 52\% and 45\% of adversaries have an accuracy of 0\% for GMF on Foursquare, Gowalla and Movielens, respectively. In contrast, The percentage of adversaries with null accuracy in FedRecs with no defense is always lower than 5\%.  This indicates a substantially larger number of adversaries unable to identify any member of the target community in GossipRecs. Similar trends can be observed for the PRME model. 

\begin{table}[!]
\centering
\caption{Summary of Attack results in the Gossip Settings. Accuracy upper bound over all configurations is 81\% and 72\% for Rand-Gossip and Pers-Gossip, respectively.\label{tab:gl}}
\tabcolsep=0.04cm
\begin{tabular}{|c|c|c|c|c|}
\hline
\begin{tabular}[c]{@{}c@{}}Gossip \\ Protocol\end{tabular} &
  \begin{tabular}[c]{@{}c@{}}Dataset\\ (Random bound)\end{tabular} &
  Model &
  \begin{tabular}[c]{@{}c@{}}Max \\ AAC \%\end{tabular} &
  \begin{tabular}[c]{@{}c@{}}Best 10 \%\\ Max AAC \%\end{tabular} \\ \hline
\multirow{5}{*}{Rand-Gossip} & \begin{tabular}[c]{@{}c@{}}Movielens\\ (5.3)\end{tabular}                   & GMF  & 14.6 & 32 \\ \cline{2-5} 
                             & \multirow{2}{*}{\begin{tabular}[c]{@{}c@{}}Foursquare\\ (3.6)\end{tabular}} & GMF  & 7.12 & 10 \\ \cline{3-5} 
                             &                                                                             & PRME & 5.8  & 12 \\ \cline{2-5} 
                             & \multirow{2}{*}{\begin{tabular}[c]{@{}c@{}}Gowalla\\ (5)\end{tabular}}      & GMF  & 11   & 20 \\ \cline{3-5} 
                             &                                                                             & PRME & 5.8  & 16 \\ \hline
\multirow{5}{*}{Pers-Gossip} & \begin{tabular}[c]{@{}c@{}}Movielens\\ (5.3)\end{tabular}                                                                   & GMF  & 14.6 & 28 \\ \cline{2-5} 
                             & \multirow{2}{*}{\begin{tabular}[c]{@{}c@{}}Foursquare\\ (3.6)\end{tabular}}                                                 & GMF  & 8    & 14 \\ \cline{3-5} 
                             &                                                                             & PRME & 5.8  & 12 \\ \cline{2-5} 
                             & \multirow{2}{*}{\begin{tabular}[c]{@{}c@{}}Gowalla\\ (5)\end{tabular}}                                                    & GMF  & 12.2 & 16 \\ \cline{3-5} 
                             &                                                                             & PRME & 6.6  & 9  \\ \hline
\end{tabular}%
\end{table}
\subsection{Impact of Personalization on CIA}\label{subsec:Q3}
We can further compare the privacy leakage between Rand-Gossip and Pers-Gossip. Although the differences in Max ACC observed in Table~\ref{tab:gl}  are relatively similar, it is still notable that the personalized setting tends to exhibit higher attack accuracy, especially considering it has a lower accuracy bound (\ie, 72\% versus 81\% as attackers meet less victims in this protocol since it is less peer exploratory). We conclude that an adversary can indeed take benefit from the personalized models, but as CIA depends on the exploration of the user space (\eg, accuracy bound), this impact is largely reduced by the more restrained peer sampling protocol of Pers-Gossip~\textbf{(RQ3)}. 

\subsection{Impact of colluders on CIA}\label{subsec:Q4}
Table~\ref{tab:colluders} presents the Max AAC for 5\%, 10\%, and 20\% colluders in a Rand-Gossip setting. The accuracy bound increases from 40.96\% to 100\% as the adversary's view size expands, aligning with the accuracy bound in FedRecs. However, the maximum average accuracy for 20\% colluders reaches only 45\%, significantly lower than in FL (45\% vs. 57.3\% average and 60.3\% vs. 76\% for the best 10\% AAC). This difference persists despite both settings sharing a 100\% accuracy bound, due to the colluders receiving models at different stages (\ie, temporality). Furthermore, achieving a global view would require them controlling at least 812 of 1083 users in a 3-regular graph, which is impractical. Thus, GossipRecs demonstrate greater resilience to CIA under realistic colluder numbers\textbf{(RQ4)}.

\begin{table}
\centering
\caption{
Effects of collusion in GL (Rand-Gossip) with GMF on MovieLens. The random bound is at  5.3\% and the Accuracy upper bound is at 100\%.\label{tab:colluders}}
\tabcolsep=0.11cm
\begin{tabular}{|c|cc|}
\hline
\multirow{2}{*}{Setting} & \multicolumn{2}{c|}{Metric}                                       \\ \cline{2-3} 
                         & \multicolumn{1}{c|}{Max AAC \%} & \multicolumn{1}{l|}{Best 10\% AAC \%} \\ \hline
Single Adversary         & \multicolumn{1}{c|}{14.6}        &          32                          \\ \hline
5\% Colluders           & \multicolumn{1}{c|}{24.8}          &   40                                 \\ \hline
10\% Colluders           & \multicolumn{1}{c|}{31}        
&          56                           \\ \hline
20\% Colluders           & \multicolumn{1}{c|}{45}        &  60.3                                  \\ \hline
\end{tabular}
\end{table}

% \begin{figure}
%     \centering
% \includegraphics{Plots/colluders.pdf}
% \caption{Effects of collusion in a gossip-based setting with Rand-Gossip on Foursquare. Note that the attack accuracy bound~(Section~\ref{subsec:protocols}) increases from $\approx$ 41\% (single adversary) to 100\% (10\% and 20\% colluders).}\label{fig:glc}
% \end{figure}

% \begin{figure}[!htbp]
%     \centering
% \includegraphics{submission-template/Plots/CDF_Colluders.pdf}
% \caption{
% Cumulative Distribution Function of the attack accuracy of all possibles attackers at the best average accuracy round: a comparison between a colluding and none-colluding settings with both full models and Share less policies on GMF for Foursquare.}\label{fig:cdfcolluders}
% \end{figure}

\section{Defense Evaluation}\label{sec:defenseeval}
 In this section, we investigate the ability of the defense strategies to mitigate CIA along with their impact on utility. More specifically, we aim at answering the following questions:

\begin{itemize}
\item \textbf{RQ5}: What is the impact of using the Share less strategy on the performance of CIA in FedRecs, GossipRecs and GossipRecs with colluders? 
\item \textbf{RQ6}: What is the Privacy/Utility trade-off of the Share less strategy?
\item \textbf{RQ7}: What is the impact of using DP-SGD on FedRecs and GossipRecs?
\end{itemize}
We answer these questions in sections \ref{subsec:Q51}, \ref{subsec:Q52},\ref{subsec:Q53}; \ref{subsec:Q6}; and \ref{subsec:Q7},~\ref{subsec:Q8}; respectively. 

% \begin{figure}
%     \centering
%     \includegraphics[width=0.45\textwidth]{submission-template/Plots/CDF.pdf}
%     \caption{Cumulative Distribution Function of the attack accuracy of all
% possibles attackers at the best average accuracy round: a comparison between
% federated and gossip settings, on both full models and Share less policies across datasets and models.}
%     \label{fig:cdf_shareless}
% \end{figure}

\subsection{Share less strategy on FedRecs}\label{subsec:Q51}
Figures~\ref{fig:tradeoff} and \ref{fig:tradeoff_prme} display the privacy/utility trade-off for the different protocols, using GMF and PRME, respectively. Furthermore, these figures serve as evidence that Share less manages to reduce CIA's accuracy in all FedRecs. Specifically, Figure~\ref{fig:tradeoff} indicates a drop from 44.93\% to 40.08\% on Foursquare; on Gowalla, a drop from 57.3\% to 41\%; and on Movielens it decreases from 57.4\% to 39.4\%. Similar, though lesser reductions can also be noted for the PRME model. In conclusion, the Share less strategy undeniably reduces the impact of CIA but is not sufficient in FL~\textbf{(RQ5)}.

% On the other end of the spectrum, there is a noticeable improvement, with 10\%-16\% of adversaries falling below the random guess on the three datasets for GMF. This contrasts with the maximum of 10\% observed when sharing full models.
% (\ie, 18\%, 41\% and 18\%) but are still much lower than those in GossipRecs.
% More specifically, on Foursquare the accuracy decreases from 19.8\% to 16.8\% while on Gowalla it goes down from 37.2\% to 30.8\%.

% \mohamed{Si on enlève le figure par round pour shareless, faudra enlever ça}
% Also, the decrease in attack accuracy is smoother in the Share less setting, where the accuracy grows at a slower pace, taking until the round 100 to reach its peak in Foursquare and Gowalla (See~\Cref{fig:fl}), and decreases even more gradually on both datasets. We attribute this slower progression to the fact that user-independent features take significantly longer to personalize and become sensitive. However, once personalized, they are difficult to fuse as they are not aggregated as often as users-dependent features. For instance, the items features space on Gowalla has a size of 32,924, and due to sparsity, it is only partially updated, in contrast to the 718 user embeddings that are updated at each step in the full model sharing.

\subsection{Share less strategy on GossipRecs}\label{subsec:Q52}
\begin{figure}
    \centering
    \includegraphics[width=0.45\textwidth]{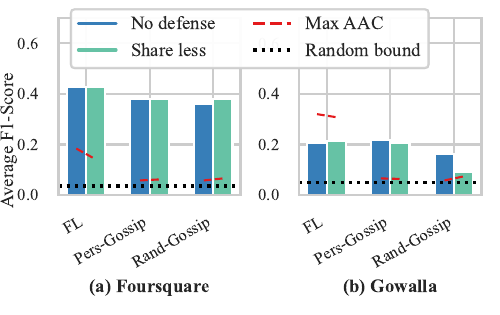}
    \caption{Attack Accuracy and F1-Score trade-off summary for the full models and Share less strategies on PRME.}
    \label{fig:tradeoff_prme}
\end{figure}

In contrast with the results observed for FedRecs, the Share less strategy appears to be largely counter-productive in the GossipRecs setting (See Figures~\ref{fig:tradeoff} and~\ref{fig:tradeoff_prme}). Indeed, we observe a surprising increase in Max
AAC (between -0.4 and +2.6) on most settings. We attribute these
results to the temporality aspects of GL. Specifically, the Share less strategy’s natural regularization effects makes models age at a slower pace, allowing for attackers to compare models coming at different rounds. In conclusion, the Share less strategy is unlikely to be efficient in mitigating CIA in GL~\textbf{(RQ5)}.

% adopting the Share less strategy does not systematically improve the gossip settings. Specifically, we observe an improvement on the GMF model for the three datasets, where an increase of of up to 10\% in the number of adversaries falling below the random bound is observe. In terms of average attack accuracy,Figure~\ref{fig:tradeoff} shows a decrease on the three datasets for Pers-Gossip, where we note an accuracy of 7.12\% 9.6\% and 10.2\% on Foursquare, Gowalla and Movielens, respectively, compared to previous averages of 8\%, 12.2\% and 14.6\%, respectively. Similar results can be observed for Rand-Gossip. There are some cases, where partial models are worst than sharing all of the models (See Figure~\ref{fig:tradeoff_prme}). As we detail in~\ref{subsec:Q6}, this is a consequence of a gain in utility observed for these models when sharing less parameters (\ie, with more personalization). The rationale here is that the better the models, the more sensitive they are. All in all, the results indicate that the Share less policy can be useful in a gossip setting where the starting level of leakage is not as significant.~\textbf{(RQ6)}. 

\subsection{Share less strategy with colluders}\label{subsec:Q53}
Table~\ref{tab:colluders_shareless} presents Max AAC for the colluding Rand-Gossip setting with the Share less strategy. Interestingly, it appears that the advantage of colluding in this setting is hardly perceptible. Specifically, we observe that 20\% of colluders are needed to achieve a 16\% MAX AAC. Moreover, an accuracy smaller by a factor of more than 2.8, compared to the same setting in a full model sharing strategy. This makes the Share less strategy more appealing in a gossip setting with colluders~\textbf{(RQ5)}.

\begin{table}
\centering
\caption{
Effects of collusion in GL (Rand-Gossip) in the Share less strategy setting. The model used here is GMF on MovieLens dataset. The random bound is at  5.3\% and the Accuracy upper bound is at 100\%.\label{tab:colluders_shareless}}
\tabcolsep=0.11cm
\begin{tabular}{|c|cc|}
\hline
\multirow{2}{*}{Setting} & \multicolumn{2}{c|}{Metric}                                       \\ \cline{2-3} 
                         & \multicolumn{1}{c|}{Max AAC \%} & \multicolumn{1}{l|}{Best 10\% AAC \%} \\ \hline
Single Adversary         & \multicolumn{1}{c|}{14}        &          29                          \\ \hline
5\% Colluders           & \multicolumn{1}{c|}{11.3}          & 19                                   \\ \hline
10\% Colluders           & \multicolumn{1}{c|}{12.8}        
&          20                           \\ \hline
20\% Colluders           & \multicolumn{1}{c|}{16}        &    28                                \\ \hline
\end{tabular}
\end{table}
\subsection{Privacy/Utility Trade-off of Share less}
\label{subsec:Q6}
\begin{figure}[h]
    \centering
    \includegraphics[width=0.45\textwidth]{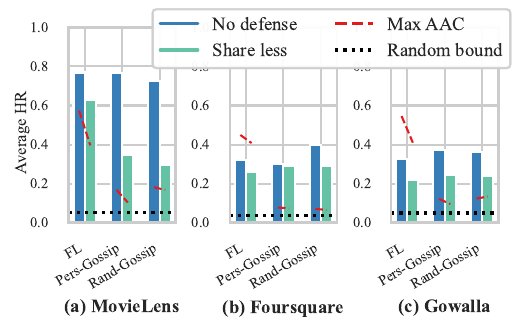}
    \caption{Attack Accuracy and Hit Ratio@20 trade-off summary for the full models and Share less strategies on GMF.}
    \label{fig:tradeoff}
\end{figure}

For GMF, utility consistently declines across all configurations, with losses ranging from 8.5\% to 16\% for FedRecs and 1.06\% to 46.5\% for GossipRecs (See Figure~\ref{fig:tradeoff}). This is due to user embeddings failing to fully offset the semantic loss caused by item embedding regularization. In contrast, PRME shows no systematic utility decrease; in some cases, it even improves slightly (\eg, a 4\% F1-Score increase on Foursquare). This may stem from the extra personalization achieved by focusing on local preferences with Share less strategy. Privacy-wise, the Share less strategy significantly benefits FedRecs, but drawing firm conclusions for GossipRecs is difficult since they already approach the random bound without defense\textbf{ (RQ6)}.

% Notably, this impact is more pronounced in the case of GossipRecs due to the absence of user embedding sharing, which prevents the unification of the latent user space. This distinction becomes particularly prominent when compared to FL, where the Share less and full model sharing approaches are equivalent, except for the regularization component. 

\begin{figure}
    \centering
    \includegraphics{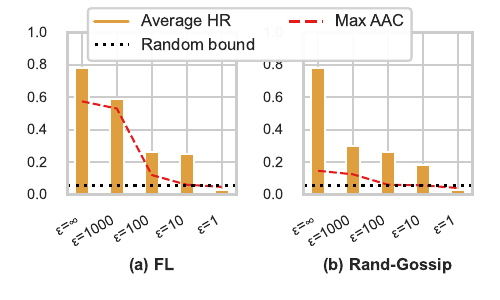}
    \caption{Average utility and empirical privacy trade-off on Movielens under DP-SGD  with different values of privacy budget $\epsilon$, $\delta = 1e^{-6}$ and clipping = 2. Utility is measured in Hit Ratio.}
    \label{fig:dp}
\end{figure}

\subsection{DP-SGD on FedRecs}\label{subsec:Q7}
In Figure~\ref{fig:dp}, left side, we directly depict the results in terms of privacy/utility trade-off of DP-SGD in the FedRecs setting. This  is particularly important, because as in many use cases of DP for complex high dimensional models, it is difficult to achieve proper utility results with meaningful privacy budget ($\epsilon$). Indeed, even with $\epsilon$ as high as 100 which would be equivalent to no actual (formal) privacy guarantee the utility is divided by a factor of 2.92. This drop-off is drastically more significant than the one observed in the Share less strategy, which in comparison, offers a much satisfying utility for a reasonable empirical privacy~\textbf{(RQ7)}. 

% This is due to the fact that DP protects against any adversary - even unrealistic ones - which means that in order to have a meaningful bound, too much noise needs to be added in our use case. To achieve $\epsilon=1$, we would need to completely destroy the utility. In any case, We still advocate for the usage of noise during model training and to control its amount with the attack's empirical accuracy. However, studying it with the lens of DP would not be useful since no meaningful bound can be found. This is why, there is a line of research to design relaxations of DP to achieve better privacy/utility trade-off but this out of the scope of this paper}. 

\subsection{DP-SGD on GossipRecs}\label{subsec:Q8}
In Figure~\ref{fig:dp}, right side, we illustrate the privacy/utility trade-off when applying DP-SGD in the Rand-gossip setting. The patterns observed are consistent with those seen in FL. More specifically, we notice that even when using higher values of $\epsilon$ that do not provide viable privacy guarantees, the utility metric is still significantly impacted. For instance, with $\epsilon = 1000$, the utility is at 30\%, which is 2.40 times lower than what can be reached without noise. Notably, even at this utility level, the attack performs considerably better than the random guess (11.2\% versus 5.3\%)~\textbf{(RQ7)}.

\section{Attack variations and parameter sensitivity analysis}\label{subsec:sensitivity}
In this section, our focus is directed towards exploring the effects of two critical parameters of CIA, namely, the momentum  (See Section~\ref{subsec:momentum}) and the size of the communities K (See Section~\ref{subsec:kvalues}). Subsequently, we delve into an investigation of attack strategies that could serve as potential proxies for CIA and perform a complexity analysis with these latter~(See Section~\ref{subsec:attacks}).

% \subsection{Different Momentum values}\label{subsec:valmomentum}
% In this section, we investigate the impact of the value of momentum on the average attack accuracy. We also consider a setting without momentum (\ie, momentum = 0), that is, a setting where the adversary launches an attack at round $t$ strictly based on models received at round $t$. Figure~\ref{fig:momentum_values} illustrates the obtained results. We observe that without momentum, the attack accuracy drops from an initial 54\% at the first round to 38\% in just 20 rounds. This decrease is continuous over rounds and can be attributed to the fact that models age differently, hence become less and less comparable overtime. In contrast, with momentum, the variance is reduced and the average attack accuracy is stable. 
% \begin{figure}
%     \centering
%     \includegraphics[width=0.45\textwidth]{submission-template/Plots/momentum_values.pdf}
%     \caption{Average Attack Accuracy evolution over rounds for different values of momentum. Dataset used is Gowalla with GMF.}
%     \label{fig:momentum_values}
% \end{figure}

\subsection{Impact of Momentum on the colluding setting}\label{subsec:momentum}
\begin{table}[!h]
\centering
\caption{Max AAC with and without momentum with various ratios of colluders in GL on MovieLens with GMF.}\label{table:colludersmomentum}
\begin{tabular}{|c|c|c|c|}
\hline
Setting  & 5\% Colluders & 10\% Colluders & 20\% Colluders \\ \hline
$\beta= 0$ & 18.4 & 21.8 & 17.6 \\ \hline
$\beta= 0.99$ & 24.8 & 31 & 45 \\ \hline
\end{tabular}
\end{table}

Table~\ref{table:colludersmomentum} illustrates the comparison of the colluding setting with and without momentum. It is easy to see that without momentum, the usefulness of colluding is largely lost (\ie, 17.6\% versus 45\% for 20\% colluders) as the colluders acquire models that are largely heterogeneous, which undermines a comparison based attack. In fact, the accuracy of the colluding setting without momentum seems bounded, that is, increasing the number of colluders beyond a certain point does not benefit the adversary. This makes the variance reduction feature of the momentum highly important and perceptible. 

\subsection{Different K values}\label{subsec:kvalues}

\begin{table}[!h]
\centering
\caption{Impact of privacy risk assessment parameter K on the Max AAC.\label{tab:k}}
\begin{tabular}{|c|c|c|c|c|c|}
\hline
Setting                & K=10 & K=20 & K=40 & K=50 & K=100 \\ \hline
Full models  & 49   & 51   & 55   & 57   & 38    \\ \hline
Share less   & 31   & 32   & 37   & 40   & 19    \\ \hline
Random Guess \%          & 1    & 2    & 4    & 5    & 10    \\ \hline
\end{tabular}%
\end{table}

Table~\ref{tab:k} showcases CIA Max AAC across different K values within FL for both model sharing strategies. Notably, we observe that changes in K do not notably influence accuracy for smaller K values. This is particularly appealing for the adversary as smaller communities often hold more relevance. As shown by the random guess value, these consistent accuracies are obtained in harder setups. We note that beyond some K value (\eg, 100) the accuracy drops, which is due to the ground truth itself as the notion of community is lost.

% \begin{figure}
% \centering
%     \includegraphics{Plots/GL-DP.pdf}
%     \caption{Average utility and empirical privacy trade-off of Rand-Gossip GMF on Movielens under DP-SGD with different values of Epsilon, $\delta = 1e^{-6}$ and clipping = 2. Utility is measured in Hit Ratio at rank S=20.}
%     \label{fig:gldp}
% \end{figure}

% \subsubsection{Extensive Analysis}
% Peer-Samppling, Spreading time, Neighbourhood size

\subsection{Detecting communities using other machine learning attacks as proxies}\label{subsec:attacks}
In this section, we study if the outputs of  pre-existing privacy attacks, namely, membership inference attacks~(MIAs) and attribute inference attacks~(AIAs), could be leveraged as proxies to conduct a Community Inference attack.
\subsubsection{Membership inference attacks as a proxy for CIA}
MIAs determine whether a given data point was part of a model's training set of a model, requiring adversaries to have access to data points likely to belong to the training set of the victim or methods to sample those candidates. In practice, an adversary can use MIA to infer the number of items in $\Dtarget$ that likely belong to the victims' training set, returning users with the most training items in $\Dtarget$ (\ie., community members). However, this approach is unsound for two reasons: primarily, MIA fundamentally differs from CIA in the fact that it detects similarities between $\Dtarget$ and training sets (\ie, memorization phenomenon), while CIA does not require $\Dtarget$ to have common items with training sets of users and instead relies on model generalization. Additionally, strong MIAs often require the costly training of shadow models~\cite{DBLP:conf/sp/CarliniCN0TT22}. 

To compare MIA and CIA, we leverage a low-cost entropy-based MIA~\cite{song2021systematic}. This attack classifies data points based on a threshold $\rho$ over the entropy of model loss. 
Table~\ref{tab:mia} demonstrates that used as a proxy for CIA, this attack it achieves a lower Max AAC, even for lower threshold values, where the training sets found by MIA have more confidence (\ie, 36\% versus 57\%). The reason for this difference in performance is that i) Inferring training membership is harder than inferring if an item aligns with a victim's preferences and ii) As explained earlier, MIA tests whether a given item has been used in the train set of the victim, while CIA tests whether the item is likely to be enjoyed by the victim.  Section~\ref{sec:complexity} includes this attack in our complexity analysis.

% Finally, using an error-prone attack like MIA~\cite{carlini2021extracting} to search for precise training data leads to a domino effect of even more errors in detecting communities.  

\begin{table}[h!]
\centering
\caption{MIA's usage as a privacy assessment metric for Community Inference in FL for GMF with Movielens.}\label{tab:mia}
\resizebox{0.45\textwidth}{!}{%
\begin{tabular}{|c|c|c|c|c|c|}
\hline
Attack & \multicolumn{1}{c|}{ $\rho=0.2$} & \multicolumn{1}{c|}{$\rho=0.4$} & \multicolumn{1}{c|}{$\rho=0.6$} & \multicolumn{1}{c|}{$\rho=0.8$} & $\rho=1$  \\ \hline
MIA Precision \% &
  44  &
  40  &
  36  &
  34  &
  0.31\\ \hline
MIA Max AAC \% & 36  & 36  & 35  & 34  & 32 \\ \hline
CIA Max AAC \%                     & \multicolumn{5}{c|}{57}                                                                                        \\ \hline
\end{tabular}
}
\end{table}

\subsubsection{Attribute inference attacks~(AIAs) as a proxy for CIA}
AIAs aim to infer sensitive attributes in a dataset based on the model it was trained on (\eg, gender in a recommendation use case~\cite{weinsberg2012blurme}). Considering community membership as an attribute, such attacks could detect communities. However, AIAs often require training a surrogate model, making them costly to launch. To investigate this, we implement an AIA attack similar to~\cite{weinsberg2012blurme}. 
The adversary randomly samples $N$ datasets from $\Dtarget$, representing fictive users belonging to the community. Similarly, $M$ datasets are sampled (with replacement) from $\itemset \setminus \Dtarget$, representing users outside of the community. On each
of these $N+M$ datasets, a GMF model is trained locally. This allows the adversary to collect gradients both from withing and outside the community members. These latter are used to train a binary classifier of five fully connected layers, with Relu activation function between layers and a
sigmoid output. It learns to classify users into community and none-community members. This classifier was then integrated into FL to mimic CIA for a randomly selected community. The Max AAC obtained indicate a 40\% accuracy for AIA compared to a 62\% for CIA, showing CIA is more effective in retrieving the community.
This is likely due to the challenge of effectively training a classifier based on gradients, as gradients from locally trained $N+M$ models do not approximate those from models in FL. In Section~\ref{sec:complexity}, we furthermore analyze the complexity of such an AIA attack compared to CIA, and the previously discussed MIA.

\subsection{Complexity Analysis of CIA compared to proxy attacks}\label{sec:complexity}
In this section, we present a temporal complexity analysis of the MIA and AIA attacks evaluated in this work, compared to CIA. Without loss of generality, we consider a scenario where the adversary aims to find a specific community $\Dtarget$ in a Share less scenario, which is a worst case scenario for CIA, from a cost perspective. We denote $T_M$, $I_M$, $T_C$ and $I_C$ as the training and inference times of the recommendation model and the classifier, respectively. We assume $I << T$ and expect $I_C \approx I_M$. Finally, we denote 
the size of the largest user training set as $D_{max}$. Table~\ref{tab:complexity} shows that CIA and MIA share similar complexities, as they both require training one fictive user embedding (\ie, $O(T_M)$) and performing a certain number of inferences. However, in practice, this number is higher for MIA as it needs to determine approximate training sets for all users before communities, thus, performing $\userset \cdot D_{max}$ inferences in the worst case, compared to the  $\userset \cdot |\Dtarget|$ inferences of CIA. In comparison, AIA requires $N+M$ more recommendation models training. It furthermore involves training a classifier on gradients. Given its input size (\ie, num items $\cdot$ embedding size) and its architecture, $T_C$  is at least as costly as $T_M$. We conclude that CIA is marginally more efficient than AIA and is more efficient than entropy-based MIA for
$|\Dtarget| < D_{max}$ and equivalent in the worst case. 
% Additionally, while the cost of training a classifier in AIA is not the sole overhead, and can indeed be amortised for the same community, this point is also valid for shadow models across other attacks, CIA included.

\begin{table}[]
\centering
\caption{Temporal Complexity of MIA and AIA compared to CIA.}\label{tab:complexity}
\begin{tabular}{|c|c|}
\hline
Attack & Temporal Complexity \\ \hline
CIA     &
$O(T_M)  + O(I_M * | U | * | \Dtarget |)$
\\ \hline
MIA    & 
$O(T_M) + O( I_M * | U |  * D_{max})$
\\ \hline
AIA    & 
$O(T_M * (N+M)) + O(T_C) + O(I_C * |U|)$
\\ \hline
\end{tabular}

\end{table}

% Notably, since adversaries do not receive gradients (or successive models) in GL, adapting AIA to GL is not straightforward.

\subsection{On the universality of CIA}\label{sec:universality}
In this paper, we focus on attacking recommender systems. The main reason for this is that the concept of ”community” within recommender systems is both intuitive and sensitive. Moreover, there is an increasing literature on FedRecs and GossipRecs overestimating the level of privacy they offer. 
Nonetheless, CIA can be used to attack other tasks. To illustrate this, in this section, we investigate the ability of CIA to generalize to an image classification task by leveraging the standard MNIST dataset~\cite{lecun1998mnist}. As the notion of community is not inherent to the dataset, we simulate it by creating a strongly non-iid data distribution between clients. Specifically, we consider 100 clients and assign to each client, samples of one unique class. Thereafter, we consider a community as a set of clients, whose assigned data points belong to the same class (\eg, community of clients that have samples representing the digit "1"). These clients, train in FL a one hidden layer of 100 units and achieve a final global accuracy of 87\%. The server tries to detect all of the communities of digits using CIA and the results show that the attacker achieves a 100\% accuracy (the random guess is at 10\% accuracy). This shows that CIA can be leveraged outside of recommender systems. The only conditions required are that the users' data distributions are non-iid and that groups of users share common distributions that form a community.

\section{Discussion}
\label{sec:discussion}
\textbf{On the fairness of the FL/GL comparison.}
Privacy considerations are crucial when selecting collaborative learning protocols. This paper delves into these privacy aspects. While such analysis might appear to strengthen the attack scenario for FL, it aligns with established assumptions in the literature~\cite{mrini2024privacy}. Additionally, we evaluate the degree of collusion required by a GL adversary to match the accuracy of an FL server, ensuring a fair comparison.

\textbf{On the usage of Secure Aggregation.}
Secure Aggregation (SA) is a multi-party computation protocol designed to protect individual models during aggregation, making it effective against privacy attacks like AIA and CIA. While SA is well-suited for Federated Learning (FL) due to cost-efficient cryptographic tools~\cite{bell2020secure}, adapting it to other collaborative learning protocols, such as Gossip Learning (GL), presents significant challenges. SA also struggles with personalization and Byzantine-resilient learning—key for recommender systems—since these often require access to individual models, as in approaches like GossipPers. Although encrypted models can sometimes address Byzantine resilience~\cite{so2020byzantine, velicheti2021secure}, this often incurs high communication costs, even in FL.

To address these limitations, advancing the flexibility of SA and exploring the privacy threat landscape through model-targeted attacks are critical, complementary research directions. This work focuses on the latter research direction.
\section{Related Work}
\label{sec:relatedwork}
\textbf{Privacy in federated recommendation systems.}
Zhang
et al.~\cite{zhang2021membership} introduced a membership inference attack~(MIA) that aimed to identify the participation of a specific user in the training process of a recommendation system, while the attack proposed in~\cite{yuan2023interaction} aims at inferring the membership of an interaction (\ie, user-item pair) to the training set. Zhang et al.~\cite{zhang2022comprehensive} proposed an attribute inference attack. Their observations in the centralized setting corroborates ours. However, they did not investigate the impact within GL. These works~\cite{zhang2021membership,zhang2022comprehensive} considered LDP as a defense mechanism, while  Yuan et al.~\cite{yuan2023interaction} proposed the Share less strategy evaluated in our work. Another line of work~\cite{perifanis2023fedpoirec} leverages Secure Multi-Party Computation~(SMPC) to secure mobility recommendation in FL. 
Nevertheless, such an approach suffers from an increased complexity as well as a Byzantine-resilience incompatibility, as discussed earlier. 

\textbf{Privacy in Gossip protocols} 
There have been few research efforts examining the privacy of decentralized learning protocols~\cite{bellet2020started, jin2023privacy}. Nevertheless, the closest works to ours are the works of~\cite{mrini2024privacy, pasquini2022privacy}, which assessed the privacy of decentralized learning through various attacks. While they did not specifically explore Community Inference, the main distinction with our work lies mostly in the setting (\ie, fixed graphs with synchronous communication). Our work suggests that the inherent privacy of gossip stems primarily from its randomness and dynamics.

\section{Conclusion}
\label{sec:conclusion}
This paper evaluates the vulnerability of Federated Learning (FL) and Gossip Learning (GL)-based recommender systems to a novel, low-cost Community Inference Attack (CIA). Unlike membership and attribute inference attacks, CIA leverages comparisons between models rather than analyzing individual models, making it particularly cost efficient. It detects communities based on shared tastes, revealing that even locally stored data can leak user similarity through model exchanges. The leakage is worse in FL compared to GL. Among the tested defenses, Share less outperformed DP by achieving a better privacy-utility trade-off. Future work will explore new defenses to strengthen FedRecs and GossipRecs against attacks like CIA.

% \input{submission-template/reviews}

%-------------------------------------------------------------------------------
% \bibliographystyle{plain}
\bibliographystyle{IEEEtran}
\bibliography{main}

% \clearpage
% \input{submission-template/reviews}

\clearpage

%%%%%%%%%%%%%%%%%%%%%%%%%%%%%%%%%%%%%%%%%%%%%%%%%%%%%%%%%%%%%%%%%%%%%%%%%%%%%%%%
\end{document}